\documentclass[12pt,onecolumn,final]{IEEEtran}
\usepackage{amsmath,amssymb}
\usepackage{danmacros}
\usepackage{times}
\usepackage{mathbf-abbrevs}
\usepackage{citesort}
\usepackage{units}
\usepackage{fancyhdr}
\usepackage{amsbsy}

\usepackage[dvips]{graphicx}
\usepackage{subfigure}
\usepackage[ruled]{algorithm2e}
\providecommand{\U}[1]{\protect\rule{.1in}{.1in}}
\renewcommand{\baselinestretch}{1.8}
\newcommand{\Es}{\Ecal_{\text{se}}}
\newcommand{\EsLB}{\Ecal_{\text{se,LB}}}

\DeclareMathOperator*{\argmin}{\operatorname{argmin}}
\newcommand{\NT}{N_{\textrm{T}}}
\renewcommand{\mod}[1]{\operatorname{mod}#1}
\newcommand{\Cube}{\textsf{CUBE}}

\newcommand{\papertitle}{\LARGE{Sum Rates, Rate Allocation, and User Scheduling for Multi-User MIMO Vector Perturbation Precoding}}
\begin{document}

\title{\papertitle}
\author{\normalsize\authorblockN{{Adeel~Razi$^{1,3}$,~Daniel~J.~Ryan$^2$,~Iain~B.~Collings$^3$~and,~Jinhong~Yuan$^1$}}\\
\small\authorblockA{$^1$School of Electrical Engineering \& Telecommunications,
The University of New South Wales, \textsc{Australia} }\\
\authorblockA{$^2$Department of Electronics and Telecommunications, Norwegian University of Science and Technology, \textsc{Norway}}\\
\authorblockA{$^3$Wireless Technologies Laboratory, CSIRO ICT Centre, \textsc{Australia}}\\
\authorblockA{\small \texttt{adeel.razi@student.unsw.edu.au, danryan@ieee.org}}%
\thanks{The material in this paper appeared in part at the 2009 IEEE InternationalConference on Communications, Dresden, Germany, June 2009. The work of A. Razi was carried out while he was on paid study leave from the NED University of Engineering and Technology, Karachi, Pakistan and was also supported in part by the Wireless Technologies Laboratory, CSIRO ICT Centre, Sydney, Australia.
The work of D. J. Ryan was supported by the Research Council of Norway (Grant 171133/V30) and was carried out in part while he was with The School of Electrical and Information Engineering, The University of Sydney, Sydney, Australia.}}%
\maketitle
\thispagestyle{empty}
\begin{abstract}
This paper considers the multiuser multiple-input multiple-output (MIMO) broadcast channel.  We consider the case where the multiple transmit antennas are used to deliver independent data streams to multiple users via vector perturbation. We derive expressions for the sum rate in terms of the average energy of the precoded vector, and use this to derive a high signal-to-noise ratio (SNR) closed-form upper bound, which we show to be tight via simulation. We also propose a modification to vector perturbation where different rates can be allocated to different users. We conclude that for vector perturbation precoding most of the sum rate gains can be achieved by reducing the rate allocation problem to the user selection problem.  We then propose a low-complexity user selection algorithm that attempts to maximize the high-SNR sum rate upper bound. Simulations show that the algorithm outperforms other user selection algorithms of similar complexity.
\end{abstract}

\section{Introduction}

Multiuser multiple-input multiple-output (MIMO) technologies may be employed by cellular base stations and wireless LAN access points to transmit messages to $K$ multiple non-collocated users without resorting to increasing bandwidth or transmit power. By exploiting the richness of multipath environments, such systems are able to achieve downlink data rates that scale linearly with the number of antennas at the transmitter, as is possible with simpler point-to-point MIMO communications, \eg \cite{Vishnawath03}, \cite{Jindal05}.

An optimal sum rate achieving transmission method for the multiuser MIMO downlink is dirty-paper coding (DPC) \cite{Vishnawath03,CaireZFDPC03}. As this scheme  requires computationally infeasible random coding and binning operations, it remains a theoretical construction. Linear precoders such as channel inversion \cite{Peel05} and zero-forcing beamforming \cite{YooJSAC06} can be used for lower complexity implementations.

A promising practical transmission method with better performance than linear precoders is vector perturbation (VP) precoding  \cite{HochwaldVP05}. With VP precoding, the data vector to be transmitted is constrained to lie within a 2$K$-dimensional hypercube of side length one, and is modified by the addition of a \emph{perturbation vector} consisting of complex integers, before being passed through a channel inverting linear precoder. The addition of the perturbation vector significantly reduces the required transmit power, and can be removed completely by independent modulo operations at each receiver. The choice of the perturbation vector is an instance of the well-studied NP-hard problem of finding the closest lattice point, whereas here the lattice is determined by the channel. A common method to perform the search is the sphere-decoding algorithm \cite{Fincke85,HassibiICASSP02,Viterbo99}, as well as suboptimal lattice reduction methods.

Due to the perturbation process, the sum rate performance of vector perturbation systems is more difficult to analyze than linear precoding systems, and exact expressions for performance measures remain an outstanding problem. This is primarily due to the fact that the performance is a function of the average power of the precoded signal, $\Es$, as this determines the effective noise power at the output of each user's demodulator. It is hard to calculate $\Es$ since it is determined by a closest lattice point search. Closed-form representations of $\Es$ are not available, however some useful closed-form bounds have been derived in \cite{RyanICC08}. In \cite{HochwaldVP05} an expression that gave insight into the choice of perturbation vector is derived, but still required numerical simulation to evaluate $\Es$. A statistical physics based approach was used in \cite{Mueller07,Zaidel08} to derive $\Es$ in the limit as $N_T,K \tendsto \infty$, where $N_T$ is the number of antennas at the BS. The approach in \cite{Mueller07,Zaidel08} requires a number of assumptions, and also the results are in terms of a fixed-point integral equation, which requires numerical evaluation. Another related result was given in \cite{Taherzadeh07}, where it was shown that sub-optimal lattice reduction based sphere-encoding \cite{Windpassinger04} achieves the full-diversity order. Additionally, expressions for bit error rates, assuming $\Es$ is known, have been given in \cite{RyanTCPre08}.

To the authors' knowledge the sum rate of vector perturbation systems has not been analyzed. Other practical issues also remain open, such as how to select a subset of users from a set of available users, or how to allocate different rates to users in order to maximize the sum rate. Various user selection and rate allocation algorithms have been suggested for linear precoders such as zero-forcing \cite{YooJSAC06} and zero-forcing dirty paper coding \cite{Tu03} but not for vector perturbation systems. These three problems are the subject of this work.

In this paper, we provide an expression for the sum rate of vector perturbation systems based on the assumptions that $\Es$ is known exactly and the data to be transmitted is uniformly distributed. Then we show that in high-SNR regime, the effect of modulo operation diminishes hence it has no bearing on the sum rate performance of the system. Using this high-SNR property, we derive a lower bound to this sum rate, as well as an asymptotic closed-form high-SNR upper bound. Simulation results suggest that this upper bound is tight for transmit SNRs greater than 10 dB.

We then propose a modification to vector perturbation precoding so that different rates may be allocated to different users. We examine the problem of optimizing the rate allocation and propose a sub-optimal rate allocation algorithm, which uses the simple $\Es$ approximation derived in \cite{RyanTCPre08}. We see that the rate allocation improves the performance in the low-SNR regime. However, for the vector perturbation precoding system the sum rate may be well approximated by an on-off function. We numerically determine that this on-off function has  mutual information of at most 0.2992 bits less than the actual mutual information. Using this knowledge, we propose that the rate allocation problem can be reduced to one of user selection.

Therefore, we next turn our attention to the practical user selection algorithms. We propose a low-complexity algorithm for user selection for the vector perturbation precoding systems. Specifically, we propose a greedy algorithm which chooses users successively in order to maximize the new sum rate upper bound at high SNR. We show that the selection criterion becomes equivalent to the selection criterion used in algorithms proposed in \cite{Tu03,YooJSAC06}, but differs in the user shedding criterion. We provide simulation results that show that the sum rate of our system is very close to that achieved by an exhaustive search through all possible combinations of users, and our proposed algorithm outperforms other low-complexity algorithms \cite{Tu03,YooJSAC06}. Simulation results also show that the user selection outperforms our proposed rate allocation algorithm, and that the rate allocation algorithm provides negligible improvement if used in conjunction with user selection.

\section{System Model}\label{sec:SystemModel}

We now detail the system model. We use $(\cdot)'$ to denote matrix transpose, $(\cdot)^{\dag}$ to denote matrix conjugate transpose and $\Vol(\cdot)$ to denote the Jordan-measurable volume \cite{Shenitzer94} of a region. We use $(\cdot)^+$ to denote Moore-Penrose pseudoinverse \cite{Rao71} and also denote the set of Gaussian (complex) integers as ${\Zbb[j]}$. We use $\left\lfloor . \right\rceil$ to denote the element-wise  rounding to the nearest Gaussian integer.

We consider the downlink of a narrowband multi-user MIMO system with $\NT$ transmit antennas broadcasting to $K\leq \NT$ spatially dispersed users. Each user has a single receive antenna. The users are selected from a set of $U$ available users. Each channel realization $\Hbf \in \Cbb^{K \times \NT}$ consists of  elements $h_{k,t} \in \Cbb$ that represents the channel between the $k^\text{th}$ user and $t^\text{th}$ transmit antenna.


Given the transmitted vector $\xbf = [x_1 \ldots x_{\NT} ]' \in \Cbb^{\NT \times 1}$, the received symbol at user $k$ is given by
\begin{equation}
y_k = \hbf_k \xbf + n_k,
\end{equation}
where $n_k$ is additive white Gaussian noise with distribution of $\Ccal \Ncal(0,1)$ and $\hbf_k = [ h_{k,1} \ldots h_{k,N_T}]$. The received symbols can be combined as  $\ybf = [y_1 \ldots y_{K} ]' \in \Cbb^{K \times 1}$ to give
\begin{equation}
\ybf = \Hbf \xbf + \nbf, \label{eq:system}
\end{equation}
where $\nbf = [ n_1 \ldots n_K]'$. The transmitted vector $\xbf$ is a modified ``perturbed" and ``precoded" form of the data vector $\abf = [ a_1 \ldots a_K ] \in \Cube^K$ where $\Cube^K$ is the $K$-ary Cartesian product of the region
\begin{equation*} \label{eq:cube}
\Cube \defas \cubr{ \; a \;:\;\abs{\Real{a}} < 0.5,\;\;\abs{\Imag{a}} < 0.5}.
\end{equation*}
Clearly, $\Vol ( \Cube^K )=1$. To generate $\xbf$, the data vector $\abf$ is first perturbed and then precoded to create the sphere-encoded signal vector, $\sbf$, according to
\begin{equation}\label{eq:precode}
\sbf =  \Fbf (\abf + \pbf),
\end{equation}
where we set $\Fbf=\Hbf^+$ to be a precoding matrix and $\pbf$ is the Gaussian (complex) integer-valued perturbation vector given by
\begin{equation}\label{eq:findp}
\pbf = \argmin_{\qbf \in \Zbb[j]^{K}} \magn{ \Fbf (\abf + \qbf)}^2.
\end{equation}

Now, choosing $\pbf$ in (\ref{eq:findp}) is a well-studied NP-hard problem of finding the closest lattice point. We assume that the algorithm used to solve (\ref{eq:findp}) gives the optimal solution for the purposes of analytical tractability. An optimal approach will have complexity exponential in $K$ e.g. the sphere-decoding algorithm of \cite{Fincke85}. Some suboptimal methods of polynomial complexity may be employed for the case when $K$ is increasing, such as the lattice reduction based approach of \cite{Windpassinger04}, and the singular value decomposition based approach of \cite{AiryVTC06}. For our simulations, we used the sphere decoding algorithm proposed in \cite{Agrell2002}. 

For analytical purposes we will consider uniformly distributed inputs where $\abf$ is an i.i.d. random variable with probability distribution function $p(\abf) = \chi_{\Cube^K}(\abf)$ where $\chi(\cdot)$ is the characteristic (indicator) function.

The final step in generating $\xbf$ is to scale $\sbf$ as follows:
\begin{align}
\xbf = \sqrt{\frac{P}{\Es(\Fbf)}} \sbf, \label{eq:CnstSca}
\end{align}
where $P$ is the \emph{transmit} signal to noise ratio (SNR), and
\begin{equation}
\Es(\Fbf) \defas E_{\abf} [\magn{\sbf}^2] =
E_{\abf}  \sqbr{ \min_{\qbf \in \Zbb[j]^{K}} \magn{ \Fbf (\abf + \qbf)}^2 }
\label{eq:Ese}
\end{equation}
is the expected power of the sphere-encoded vector $\sbf$ for a channel instance (packet) $\Hbf$, where the expectation is taken over $\abf$. That is, the expected power required to transmit each packet is constant. Hence the receiver only needs to know $\Es$, which is a data independent quantity, in order to decode the received signal correctly\footnote{In a practical system, the transmitter would calculate the packet power and then scales the packet to satisfy the power constraint. If the packet is long enough, the empirical and expected values of $\Es$ will be close.}. 


At the $k$th user's receiver, the data is recovered using a modulo demodulator \cite{HochwaldVP05}
\begin{align}
\ah_k
&= \sqbr{ \sqrt{\Es(\Fbf)/P}  y_k  }_{\mod\Cube^K}  
= \sqbr{a_k + p_k + \sqrt{\Es(\Fbf)/P} n_k  }_{\mod\Cube^K} \nonumber  \\
&= \sqbr{ a_k + \eta_k }_{\mod\Cube^K} , \label{eq:system_modulo}
\end{align}
where, $a_k, p_k,$ and $n_k$ are the $k$th element of the vectors $\abf, \pbf,$ and $\nbf$ respectively and $\eta_k \defas \sqrt{\Es(\Fbf)/P} n_k $ is the effective noise for user $k$. Therefore $\Var{\eta_k} = \Es(\Fbf)/P$. The function $[.]_{\mod\Cube^K}$ denotes a modulo operation which is defined as $[.]_{\mod\Cube^K}=[.]-\left\lfloor .\right\rceil$.  This operation finds a point inside the region $\Cube^K$ if the point lies outside the region $\Cube^K$. The modulo operation is applied to the real and imaginary parts independently.

\section{Sum Rate of Vector Perturbation Precoding} \label{sec:SumrateVP}

In this section, we derive the sum rate of the VP precoding system using uniformly distributed inputs given that the value of $\Es(\Fbf)$ is known. We derive a lower bound to this sum rate which is also approached asymptotically as the transmit SNR $P \tendsto \infty$. We then derive an upper bound to the sum rate using a lower bound to $\Es(\Fbf)$ that we recently derived in \cite{RyanICC08}.

First, we derive an expression for the sum rate of the VP precoding system in terms of $\Es(\Fbf)$. Define $I(\ah_k;a_k | \Hbf,\Fbf)$ as the mutual information between $\ah_k$ and $a_k$ given channel matrix $\Hbf$ and precoding matrix $\Fbf$.
\begin{theorem}\label{th:VPcapexact}
The sum rate $R_{\text{VP}}$ of an $N_T \times K$ vector perturbation system with uniformly distributed inputs is
\begin{align}\label{eq:Rvp}
R_{\text{VP}}(\Hbf,\Fbf)
&\defas \sum_{k=1}^K I(\ah_k;a_k | \Hbf, \Fbf) \nonumber \\
&= K \log \frac{P}{K} - K \log \frac{\pi e \Es(\Fbf)}{K}  + 2K \Omega \br{\frac{\Es(\Fbf) }{2P}}
\end{align}
where
\begin{equation} \label{eq:Omegadef}
\Omega \br{\gamma} = \frac{1}{2} + \int_{-\frac{1}{2}}^{\frac{1}{2}} \sum_{s = -\infty}^\infty \tfrac{1}{\sqrt{ 2 \pi \gamma}}e^{ - \frac{\abs{\xi - s}^2  }{2\gamma}} \sqbr{\log \sum_{t = -\infty}^\infty   e^{ - \frac{\abs{\xi - t}^2}{2\gamma}}} d\xi.
\end{equation}
\begin{proof}
See Appendix I.
\end{proof}
\end{theorem}

We now discuss this result. We see that $\Es(\Fbf)$ and the function $\Omega(\gamma)$ are important terms in order to understand the sum rate for the vector perturbation system, hence we go in detail to examine these two terms one by one.

With regards to $\Es(\Fbf)$, we note that no exact analytical results have yet been obtained. Some partially numerical results concerning the value of $\Es(\Fbf)$ were presented in \cite{HochwaldVP05}. In \cite{Mueller07,Zaidel08}, using replica method of statistical physics an asymptotic result for $\Es(\Fbf)$ was derived as a coupled fixed-point representation. However, for the case of uniformly distributed inputs, we derived a lower bound in \cite{RyanICC08}, which was shown to be a good approximation for most input distributions. We will subsequently use the result of \cite{RyanICC08} to derive an asymptotic upper bound on the sum rate. 

Next, we turn to the term $\Omega(\gamma)$, where $\gamma = \Es(\Fbf)/(2P)$. The term $\Omega(\gamma)$ captures the effect of the modulo operation on the Gaussian noise. We see that, from (\ref{eq:Hxi}) in Appendix I,

\begin{equation}
\Omega(\gamma) =  \frac{1}{2} \log (2 \pi e \gamma) - H(\xi), \label{eq:Omega_xi}
\end{equation}
where, $\xi = \Real{\sqbr{\eta_k}_{\mod{\Cube}}}$. As $P \tendsto \infty$, it follows that
\begin{equation*}
\lim_{P \tendsto \infty} H(\xi) = \frac{1}{2} \log (2 \pi e \gamma)
\end{equation*}
which concurs with the intuition that the distribution of $\xi$ approaches $\Ncal(0,\xi)$, as the noise variance decreases. Applying this to (\ref{eq:Omega_xi}) gives
\begin{equation}\label{eq:omegalim0}
\lim_{\gamma \tendsto 0}  \Omega(\gamma) = 0.
\end{equation}
Moreover, since $H(\xi) \leq \frac{1}{2} \log (2 \pi e \gamma)$ and as $\frac{1}{2} \log (2 \pi e \gamma)$ is the maximum entropy for any random variable with variance $\gamma$, therefore $\Omega(\gamma) \geq 0$. As $P \tendsto 0$, the distribution of $\xi$ approaches a uniform distribution over the interval $\sqbr{-\frac{1}{2},\frac{1}{2}}$. It follows that $\lim_{P \tendsto 0} H(\xi) = 0$, and thus
\begin{equation*}
\lim_{P \tendsto 0}  \Omega(\gamma) = \frac{1}{2} \log 2 \pi e \gamma.
\end{equation*}
In summary, $ \Omega(\gamma)$ is an increasing function in $\gamma$ (and decreasing in P) with range $(0,\frac{1}{2} \log 2 \pi e \gamma)$ for $\gamma > 0$. In the high-SNR regime, $ \Omega(\gamma)$ will be small, as the effect of the modulo operation diminishes, and therefore negligible when it comes to determining the sum rate.

We now use Theorem \ref{th:VPcapexact} to derive the following useful bounds and asymptotic values of the sum rate. By noting that $\Omega(\gamma) > 0$, and approaches 0 as $P \tendsto \infty$, we have the following lower bound and asymptotic result.
\begin{corollary}
The sum rate $R_{\text{VP}}$ of an $N_T \times K$ vector perturbation system with uniformly distributed inputs satisfies the lower bound
\begin{equation}
R_{\text{VP,LB}} \defas  K \log \frac{P}{K} - K \log \frac{\pi e \Es(\Fbf)}{K`}
\end{equation}
which is approached as $P \tendsto \infty$.
\end{corollary}

Additionally, we also have the following asymptotic upper bound which we will use as a basis for the user selection algorithm in Section \ref{sec:userselection}.
\begin{corollary}\label{cor:CVP-UBhighEx}
As $P \tendsto \infty$, the sum rate $R_{\text{VP}}$ of an $N_T \times K$ vector perturbation system, employing uniformly distributed inputs and precoding matrix $\Fbf = \Hbf^+$ has the following the upper bound
\begin{equation*}
\lim_{P \tendsto \infty} R_{\text{VP}} <  K \log  \frac{P}{K} + \log \det(\Wbf)
- K \log \frac{\Gamma(K+1)^{\frac{1}{K}}e}{(K+1) }.
\end{equation*}
where $\Wbf \defas \Hbf \Hbf^{\dag}$ and $\Gamma(\cdot)$ denotes the gamma function.

\begin{proof}
First, recall from our discussion of (\ref{eq:Rvp}) in Theorem \ref{th:VPcapexact} that $\Omega(\gamma) \tendsto 0$ as $P \tendsto \infty$. Then, we substitute the lower bound on $\Es(\Fbf)$ from \cite{RyanICC08}, namely
\begin{equation}\label{eq:EsHLB}
\Es(\Fbf) \geq \EsLB(\Fbf) \defas \frac{ K \Gamma(K + 1)^{1/K}}{(K+1)\pi} \det(\Fbf^\dag \Fbf)^{1/K}
\end{equation}
into (\ref{eq:Rvp}). By noting that $\Fbf=\Hbf^+$ and therefore $\Fbf^\dag \Fbf = \Wbf^{-1}$, completes the result.
\end{proof}
\end{corollary}

\section{Rate Allocation for Vector Perturbation Precoding} \label{sec:powerallocation}

In this section we will extend the system model by taking into account the rate allocation in an attempt to further optimize the sum rates. Using a rate allocation matrix $\Lambdabf$, we derive an expression for sum rate and then discuss the performance gain yielded by the rate allocation.

We propose to decompose the channel matrix $\Hbf$ as
\begin{equation}
\Hbf=\Dbf\Vbf\Qbf,
\label{eq:decomposeH}
\end{equation}
where this decomposition in (\ref{eq:decomposeH}) is a variation of QR decomposition such that $\Dbf=$  diag$(d_{1},\ldots,d_{K})$, $\Vbf$ is lower triangular with ones on its diagonal and $\Qbf$ is a unitary matrix. Then $\Hbf^{+} =  \Qbf^{+}\Vbf^{+}\Dbf^{+}$. Instead of using $\Hbf^+$ as a precoding matrix, as was the case in Sections \ref{sec:SystemModel} and  \ref{sec:SumrateVP}, we now set $\Fbf=\Qbf^+ \Vbf^+ \Lambdabf$ to be a modified precoding matrix so as to take into account the rate allocation using $\Lambdabf=$  diag$(\lambda_{1},\ldots,\lambda_{K})$ as a rate allocation matrix. Now the Gaussian (complex) integer-valued perturbation vector $\pbf$ is given by
\begin{equation}
\pbf = \argmin_{\qbf \in \Zbb[j]^{K}} \magn{ \Vbf^+ \Lambdabf (\abf + \qbf)}^2.
\end{equation}

\begin{figure}
\begin{center}
\includegraphics[width=\columnwidth]{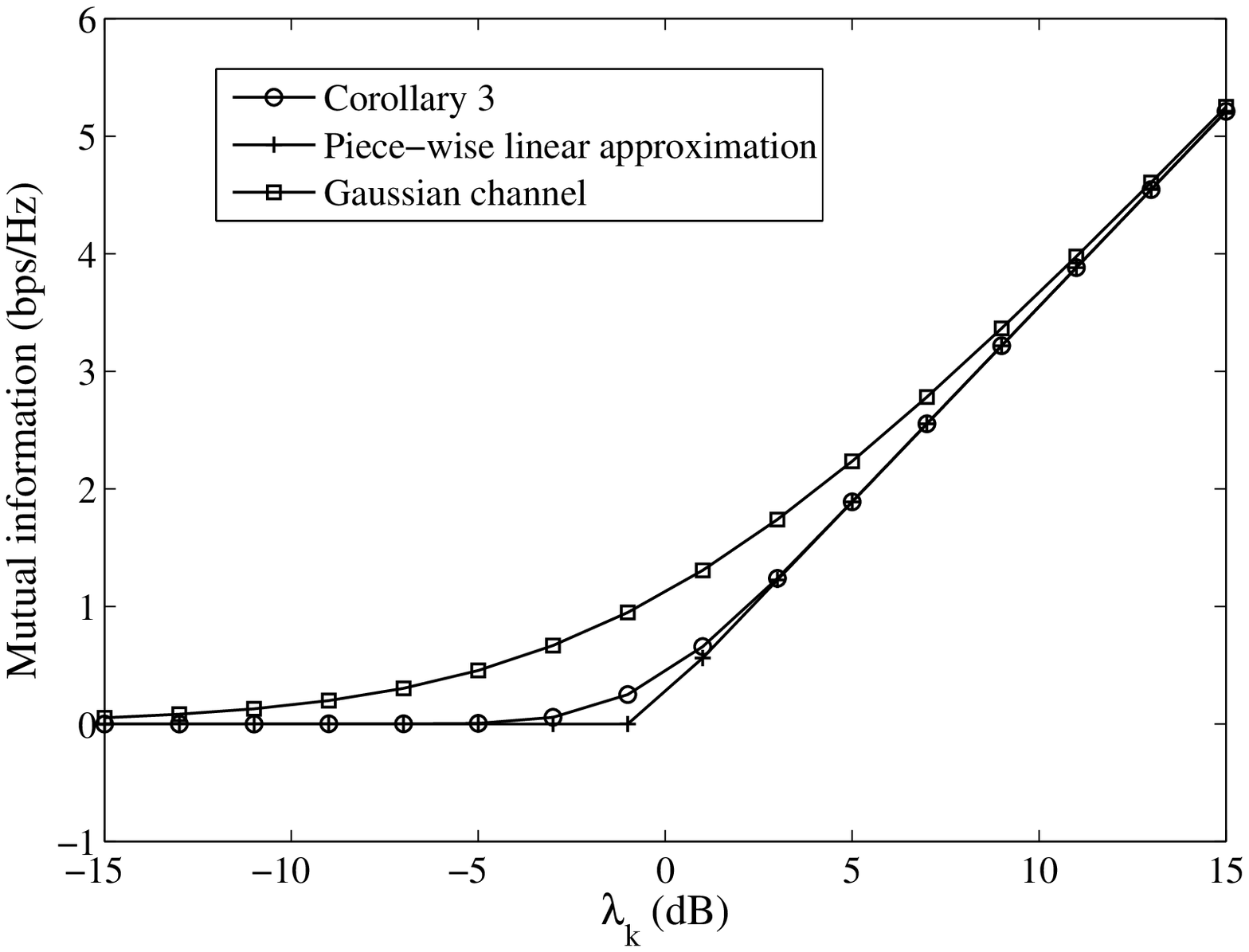}
\end{center}
\caption{Plot of mutual information (bps/Hz) versus $\lambda_k$ from Corollary \ref{cor:3}, piece-wise approximation given by (\ref{eq:Ipw}) and the Gaussian channel expression given by (\ref{eq:Iawgn}). SNR = 0 dB, $\Es = 0.1$ and $d_k = 1$.}
\label{fig:fig1}
\end{figure}

We then scale $\sbf$ to generate the transmit vector $\xbf$ as follows:
\begin{align}
\xbf = \sqrt{\frac{P}{\Es(\Fbf)}} \sbf.
\end{align}
The received signal at the $k$th user is then
\begin{equation*}
y_k =  \sqrt{\frac{P}{\Es}} d_k \lambda_k(a_k+p_k) + n_k
\end{equation*}
and the recovered data symbol at the output of the modulo demodulator of the $k$th user is given by
\begin{align}
\ah_k
&= \sqbr{ \sqrt{\frac{\Es(\Fbf)}{P \lambda_{k}^{2}d_{k}^{2}}}  y_k  }_{\mod\Cube^K}  
= \sqbr{a_k + p_k + \sqrt{\frac{\Es(\Fbf)}{P \lambda_{k}^{2}d_{k}^{2}}} n_k  }_{\mod\Cube^K} \nonumber  \\
&= \sqbr{ a_k + \eta_k }_{\mod\Cube^K}, \label{eq:new_system_modulo}
\end{align}
where $\eta_k = \sqrt{\frac{\Es}{P \lambda_{k}^{2}d_{k}^{2}}} n_k $ is the effective noise for user $k$.
\begin{corollary}\label{cor:3}
The sum rate $R_{\text{VP-RA}}$ of an $N_T \times K$ vector perturbation system with uniformly distributed inputs and precoding matrix $\Fbf = \Qbf^+ \Vbf^+ \Lambdabf$ is
\begin{align}
R_{\text{VP-RA}}(\Hbf,\Fbf)
&= \sum_{k=1}^K I(\ah_k;a_k | \Hbf, \Fbf) \nonumber\\
&= \sum_{k=1}^K \left\{ \log \frac{P\lambda_k^{2} d_{k}^{2}}{K} - \log \frac{\pi e \Es(\Fbf)}{K}  + 2 \Omega \br{\frac{\Es(\Fbf) }{2P\lambda_k^{2} d_{k}^{2}}}\right\}.
\end{align}
\begin{proof}
From (\ref{eq:new_system_modulo}) we see that $\Var{\eta_k} = \frac{\Es}{P \lambda_{k}^{2}d_{k}^{2}}$, hence by using this variance and following the steps in Theorem \ref{th:VPcapexact}, the proof is completed.
\end{proof}
\end{corollary}
We note that the choice of the optimal $\Lambdabf$ is difficult as the rate is a function of $\Es(\Fbf)$, which is an NP-hard problem to evaluate. In order to find a simple sub-optimal approach to the rate allocation problem, we first examine the mutual information function $I(\ah_k;a_k | \Es(\Fbf), d_k)$ as a function of $\lambda_k$. In Fig. \ref{fig:fig1}, we plot $I(\ah_k;a_k | \Es(\Fbf), d_k)$ as a function of $\lambda_k$ for SNR = 0 dB, $\Es(\Fbf)=0.1$ and $d_k=1$. We also plot a piece-wise linear approximation to $I(\ah_k;a_k | \Es(\Fbf), d_k)$, namely
\begin{equation}\label{eq:Ipw}
I_{\text{PW}}(\ah_k;a_k | \Es(\Fbf), d_k) = \max \cubr{0,\log \frac{P\lambda_k^{2} d_{k}^{2}}{K} - \log \frac{\pi e \Es(\Fbf)}{K}}= \max \cubr{0,\log \frac{P \lambda_{k}^{2}d_{k}^{2}}{\pi e \Es(\Fbf)}} ,
\end{equation}
as well as the mutual information of a Gaussian channel matched to have the same mutual information in the high and low SNR regimes
\begin{equation}\label{eq:Iawgn}
I_{\text{AWGN}}(\ah_k;a_k | \Es(\Fbf), d_k) = \log \br{1 + \frac{P \lambda_{k}^{2}d_{k}^{2}}{\pi e \Es(\Fbf)}}.
\end{equation}
The piece-wise linear approximation in (\ref{eq:Ipw}) is motivated by the fact that, as we showed in Section III, $\Omega(\gamma)$ approaches 0 as $P \tendsto \infty$ hence the modulo vector perturbation channel in high SNR regime is a high SNR AWGN channel. While for low SNR, it can be seen as a zero mutual information channel. Also note that expressions of the logarithmic form, as in (\ref{eq:Iawgn}), are obtained when linear precoding schemes are used with Gaussian inputs, as the received signal is also Gaussian.

We see that $I_{\text{PW}}$ is much tighter for the modulo vector perturbation channel than $I_{\text{AWGN}}$.  The maximum difference with the piece-wise approximation is at most $1$ bit for the AWGN channel and only $\sim 0.2992$ bit for the modulo vector perturbation channel. Note also that the range of $\lambda_k$ where the difference is non-negligible is much less for the piecewise approximation, which also explains why such an approximation is of less interest for linear precoding systems.

We propose to take advantage of the tightness of the piecewise lower bound to simplify the method of rate allocation. Specifically, we propose to maximize the rate allocation function
\begin{align}
R_{\text{VP,PW}}
& \defas  \sum_{k=1}^{K} \max \cubr{0,\log \frac{P \lambda_{k}^{2}d_{k}^{2}}{\pi e \Es(\Fbf)}}.
\label{eq:approxRvp}
\end{align}
From  the above we know that the maximum difference between the actual sum rate and this piece-wise approximation is at most $0.2992K \leq 0.2992\NT$ bits. To remove the difficulty in optimization imposed by the dependence on the $\Es$ function we again use the lower bound in $(\ref{eq:EsHLB})$, assuming now that the precoding matrix has
\begin{equation}\label{eq:EseLB2}
\Es(\Fbf) \geq \frac{ K \Gamma(K + 1)^{1/K}}{(K+1)\pi} \det(\Lambdabf ^2)^{1/K}
\end{equation}
as $\det(\Vbf)=1$ and $\Qbf\Qbf^\dag =1$. By inserting (\ref{eq:EseLB2}) into (\ref{eq:approxRvp}) we get
\begin{equation}\label{eq:RVPUB}
R_{\text{VP,PW}} \leq
\sum_{k=1}^{K} \max
\cubr{0, \log \frac{P}{K}- \log\frac{\Gamma (K+1)^{1/K}e}{(K+1)} + \log d_{k}^{2} + \log \lambda_{k}^{2} -\frac{1}{K} \sum_{k=1}^{K} \log \lambda_k^{2} }.
\end{equation}
The value of using (\ref{eq:EseLB2}) as an approximation has been examined in \cite{RyanTCPre08}. We now examine how the rate allocation proceeds from here. To simplify (\ref{eq:RVPUB}) we set
\begin{equation*}
c = \frac{1}{K}\sum_{k=1}^{K} \log \lambda_k^{2}
\end{equation*}
and $\log (\lambda'_k)^2 = \log \lambda_k^{2} - c$. Substituting this into (\ref{eq:RVPUB}) we obtain
\begin{equation*}
R_{\text{VP,PW}} \leq \sum_{k=1}^{K} \max \cubr{0, R_{0,k} + \log (\lambda_{k}')^2}.
\end{equation*}
where $R_{0,k} \defas \log \frac{P}{K}- \log\frac{\Gamma (K+1)^{1/K}e}{(K+1)} + \log d_{k}^{2}$.  
Now, if we place the restriction that $K$ users must be used then the sum rate is at most
\begin{align*}
R_{\text{VP,PW}} &\leq \sum_{k=1}^{K} (R_{0,k} + \log (\lambda_{k}')^2) = \sum_{k=1}^{K} R_{0,k} \nonumber \\  
&= \!K \log \frac{P}{K} + \log \det(\Wbf)- K \log \frac{\Gamma (K+1)^{1/K}e}{(K+1)} \\
&= R_{\text{VP,UB}}.
\end{align*}
Note that if $\lambda'_k$ is chosen so that  $\log d_k^{2}$ and $\log (\lambda'_k)^2$ are equal, that would imply that either all or none of the users are in the non-zero rate regime. This choice of $\lambda'_k$ corresponds to standard vector perturbation as outlined in Section \ref{sec:SystemModel}.

We see that by making this piece-wise linear approximation to the mutual information, and the use of the $\Es$ approximation, the best sum-rate obtainable due to rate allocation is approached by simply selecting users so as to maximize $R_{\text{VP,UB}}$.  To summarize, as a consequence of the modulo vector perturbation channel for a particular user being effectively a high SNR AWGN channel in the high SNR regime, and a zero mutual information channel in the low SNR regime, the difference between an on-off assumption and the modulo vector perturbation channel (0.2992 bit) is much less than the difference between the on-off assumption and the AWGN assumption ($\leq 1$ bit, and for a much greater range of gains). Consequently, we would expect that, to approach the maximum sum rate it is sufficient to select the users that will maximize the high-SNR sum rate upper bound given by Corollary \ref{cor:CVP-UBhighEx}. Moreover, it is sufficient to use the standard channel inverse precoding matrix to achieve this rate.

\section{User Selection Algorithm} \label{sec:userselection}

We now turn to the user selection, both as a rate allocation algorithm, and for use in scenarios when the number of potential users $U$ is greater than the number of transmit antennas. We propose an algorithm which we refer to as greedy rate maximization (GRM) for user scheduling for vector perturbation precoding. GRM is a low-complexity scheme, which can be considered a \emph{greedy} algorithm to maximize the capacity upper bound of Corollary \ref{cor:CVP-UBhighEx}. It turns out that the criteria for selecting users is similar to that used for zero-forcing dirty-paper coding in \cite{Tu03}, and modified for zero-forcing beamforming in \cite{YooJSAC06}. We discuss the differences in the algorithms, in terms of shedding users and terminating the user selection process. It is to be noted that our proposed greedy algorithm focus on maximizing the sum rate but in doing so fairness among the users is not guaranteed.  

The user selection algorithm we propose is as follows. Denote $\Scal$ as the set of users that have been selected, the cardinality of $\Scal$ is $K=|\Scal|$, and $\Ucal$ as the set of users who have not been selected or removed from consideration. For the selected users $\Scal$ we denote $\Hbf(\Scal)$ as the channel matrix constructed from these users, and $\Wbf(\Scal) = \Hbf(\Scal)\Hbf(\Scal)^\dag$. The algorithm we propose here maximize the high-SNR upper bound of Corollary \ref{cor:CVP-UBhighEx} by maximizing $\det(\Wbf(\Scal))$. From (\ref{eq:EsHLB}) we note that maximizing $\det(\Wbf(\Scal))$ is actually equivalent of minimizing $\Es(\Fbf)$. The algorithm is as follows:  
\begin{enumerate}
\item
Initialize the set of selected vectors $\Scal = \emptyset$, and set $\Ucal$ to the set of all users.
\item
Calculate $\det(\Wbf( \Scal \cup u))$ for all users $u \in \Ucal$. Determine $u_{\text{max}}$, the user that maximizes $\det(\Wbf( \Scal \cup u))$.
\item
Remove from $\Ucal$ all those users such that $R_{\text{VP}}$ would be reduced if they were to be added to $\Scal$. Precisely, remove user $u$ if
\begin{equation}\label{eq:step3}
\frac{\det(\Wbf( \Scal \cup u))}{\det(\Wbf( \Scal))} < \frac{e (K+1)^{2K+1}}{ P K^K (K+2)^{K+1}}
\end{equation}
and $K>1$. (We will provide a low complexity way for calculating the left hand side of this equation.)
\item
If $\Ucal$ is non-empty, add user $u_{\text{max}}$ to $\Scal$ and remove it from $\Ucal$, and return to step 2.
\item
If $\Ucal$ is empty or $K=N_T$, terminate the algorithm.
\end{enumerate}

We now compare the operations performed by GRM with Greedy-ZF \cite{Tu03} and semi-orthogonal user selection (SUS) \cite{YooJSAC06}. First, we show that the metric $\det(\Wbf( \Scal \cup u))$ in Step 2 above that determines the users to be picked, is equivalent to that used in Greedy-ZF and SUS. Thus, we show that Greedy-ZF and SUS algorithms can essentially be viewed as greedy determinant maximization algorithms. Therefore, the difference between the algorithms boils down to how the users are removed from $\Ucal$ to improve the complexity.

To show the equivalence of the choice of the next user to add to $\Scal$, we note that if we append a user $u$ with channel vector $\hbf_u$ to a set $\Scal$, and employ the block matrix determinant formula to $\det(\Wbf(\Scal \cup u))$ we obtain
\begin{align}
\det(\Wbf(\Scal \cup u))
&= \det \br{ \sqbr{ \begin{matrix} \Hbf(\Scal)\Hbf(\Scal)^\dag & \Hbf(\Scal)\hbf_u^\dag \\ \hbf_u \Hbf(\Scal)\dag & \hbf_u \hbf_u^\dag \end{matrix} }} \nonumber\\
&= \det(\Wbf(\Scal)) \magn{ \hbf_u (\Ibf - \Pbf(\Scal))}^2, \label{eq:detupdate}
\end{align}
where $\Pbf(\Scal) = \Hbf(\Scal) (\Hbf(\Scal) \Hbf(\Scal)^\dag)^{-1} \Hbf(\Scal)^\dag$ is a projection matrix for the subspace spanned by  $\Hbf(\Scal)$, which we denote $\Hcal(\Scal) \subset \Cbb^{\NT \times \NT}$. The matrix $\Ibf - \Pbf(\Scal)$ is the projection matrix for the nullspace of $\Hcal(\Scal)$. It follows from (\ref{eq:detupdate}) that the choice of user in $\Ucal$ that maximizes the determinant given $\Hbf(\Scal)$, is the user with channel vector $\hbf_u$ that has the largest component in the nullspace of $\Hcal(\Scal)$.

It is worthwhile to note that the condition given above is same as that specified by the Greedy-ZF and SUS algorithms. However, the motivations behind these other algorithms are slightly different, as the users are chosen to maximize the individual user gains in order to maximize the sum rate. In GRM we attempt to maximize the sum rate by minimizing the transmit power scaling $\Es$ via maximizing $\det(\Wbf)$. However, by noting this similarity, we are able to take advantage of the lower complexity method in \cite{YooJSAC06} to calculate the component of channel vectors orthogonal to $\Hcal(\Scal)$. That is, instead of calculating  $\hbf_u (\Ibf - \Pbf(\Scal))$, we calculate
\begin{equation}\label{eq:gk}
\hbf_u (\Ibf - \Pbf(\Scal)) = \gbf_u \defas \hbf_u  \br{ \Ibf - \sum_{s \in \Scal} \frac{ \gbf_s^* \gbf_s}{\magn{\gbf_s}^2}},
\end{equation}
where $\gbf_s$ is the value of $\gbf_u$ calculated in the previous iterations of the algorithm. Note that this makes $\gbf_s$ an orthogonal set of vectors, and that each $\gbf_u$ is also orthogonal to these vectors.  Therefore, we propose that Step 2 of the algorithm is performed by choosing the user with the greatest value of $\magn{\gbf_u}^2$, thus avoiding the calculation of determinants.

We see that $\magn{\gbf_u}^2$ can also be used for user shedding in Step 3 of the algorithm, as $\magn{\gbf_u}^2 = \det(\Wbf( \Scal \cup u))/\det(\Wbf( \Scal))$.  Note here that as $K$ increases, $\magn{\gbf_u}$ is non-increasing, and the right-hand side of (\ref{eq:step3}) is increasing. It follows that we can remove user $u$ from $\Ucal$, as it will always decrease the rate upper bound.  As we will see in the next section, this user shedding reduces the complexity of the algorithm, and results in a better sum rate performance than other algorithms.

Note that Greedy-ZF does not perform user shedding, while the SUS algorithm performs user shedding based on only keeping those vectors that are semi-orthogonal to the most recent vector added to $\Scal$. Specifically, all users satisfying
\begin{equation}\label{eq:cos2}
\cos^2 \theta (\gbf_s,\hbf_u) \defas \frac{\abs{\hbf_u \gbf_s^*}^2 }{ \magn{\hbf_u}^2 \magn{\gbf_s}^2 } > \alpha^2
\end{equation}
are removed, where $\alpha$ is a parameter in the interval $[0,1]$. Note that the optimal value of $\alpha$ for a specific antenna/user configuration and channel distribution/SNR can only be determined via simulation. This in contrast to our proposed GRM scheme, which only requires knowledge of $P$, rather than the full channel statistics.

As demonstrated in the next section, the run-time complexity of GRM, Greedy-ZF and SUS is similar. Note that SUS requires further calculation of  (\ref{eq:cos2}) as part of its user shedding calculations, thus making it more complex for the same size $\Ucal$ than our proposed GRM algorithm.

\section{Simulation Results}

\begin{figure}
\begin{center}
\includegraphics[width=\columnwidth, height=150mm]{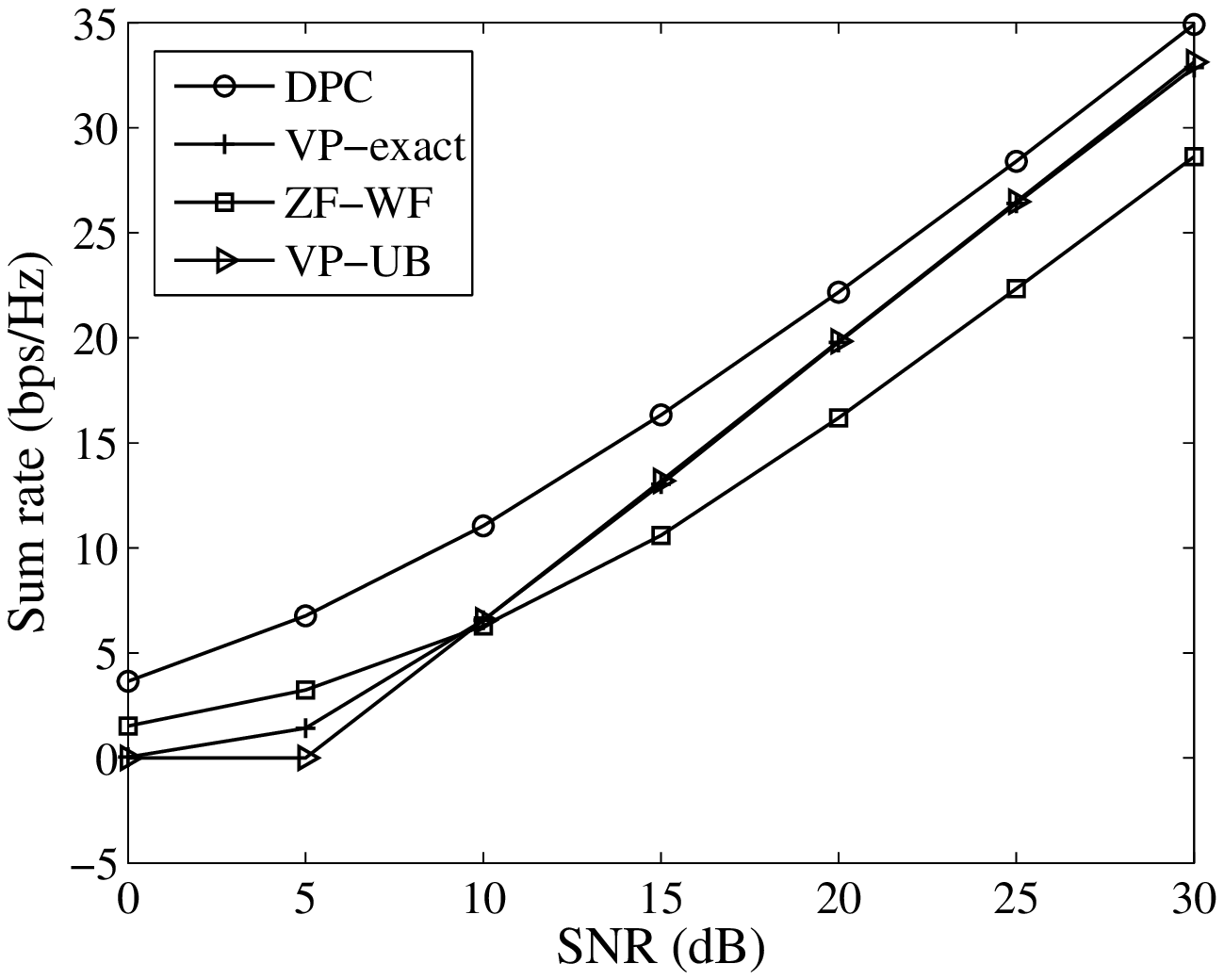}
\end{center}
\caption{Plot of sum rate (bps/Hz) versus  SNR (dB) for DPC, VP-exact, VP upper bound and zero-forcing with waterfilling (ZF-WF). $U = K = N_{T} = 4$.}
\label{fig:fig2}
\end{figure}

\begin{figure}
\begin{center}
\includegraphics[width=\columnwidth, height=150mm]{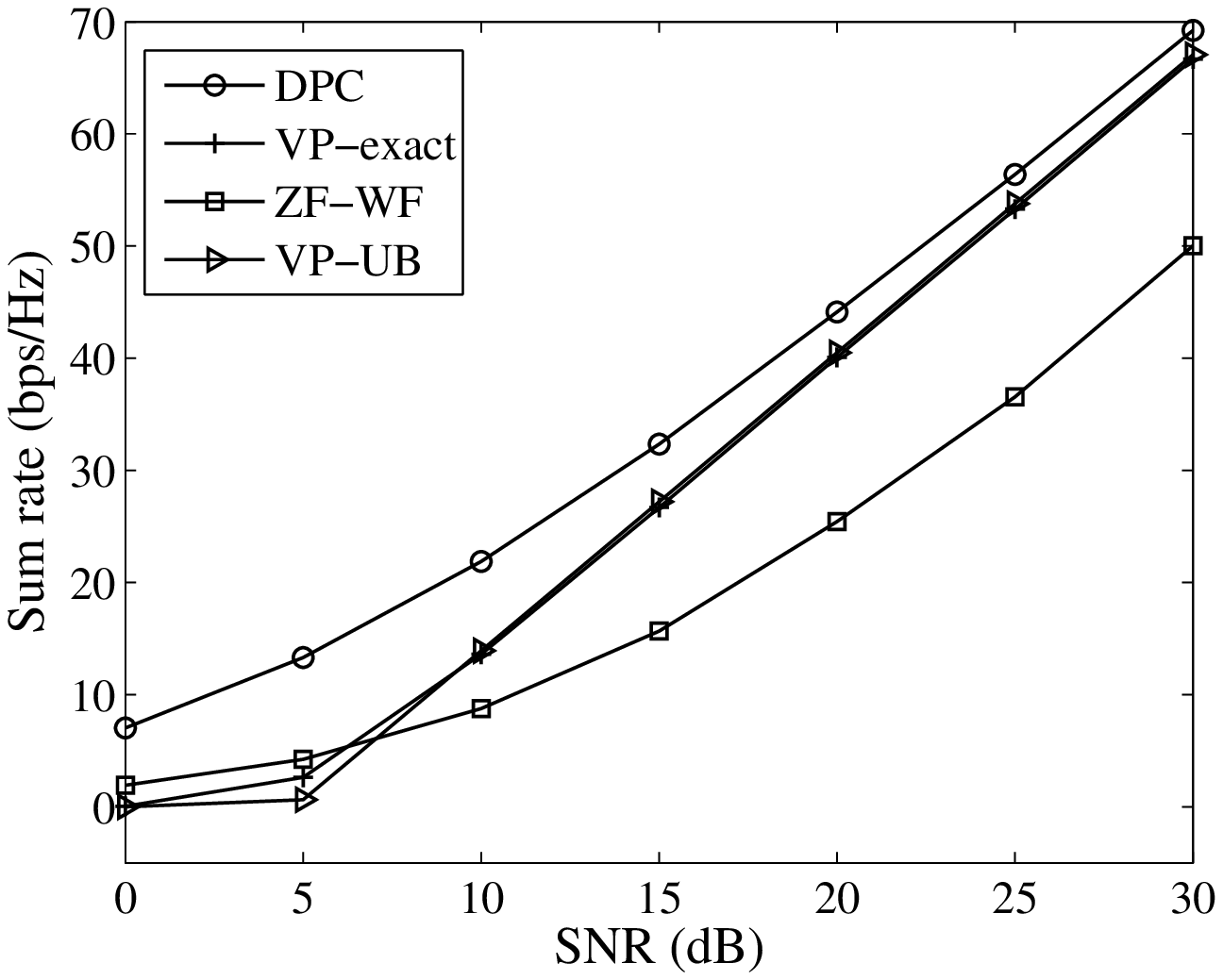}
\end{center}
\caption{Plot of sum rate (bps/Hz) versus  SNR (dB) for DPC, VP-exact, VP upper bound and zero-forcing with waterfilling (ZF-WF). $U = K = N_{T}=8$.}
\label{fig:fig3}
\end{figure}

In this section we present simulation results for sum rate performance of VP with and without user scheduling. In Figs. \ref{fig:fig2} and \ref{fig:fig3}, we consider a system with $N_{T}$ = $U$ = $K$ = 4 and 8 respectively. We plot the exact sum rate of VP precoding given by Theorem \ref{th:VPcapexact}, denoted VP-exact, where $\Es$ is generated by using Monte Carlo simulations. We also plot the high SNR upper bound for VP which is $\max\left\{0,R_{\text{VP-UB}}\right\}$ where, $R_{\text{VP-UB}}$ is given by Corollary \ref{cor:CVP-UBhighEx}. For comparison purpose, we include the plots for DPC and ZF-WF \cite{YooJSAC06}. We used 1000 independent channel realizations to obtain these plots. The plot shows that VP-exact is outperforming ZF-WF, although at low SNR ZF-WF is better due to waterfilling. We also note that the high SNR upper bound for VP is tight for SNRs greater than 10 dB.

\begin{figure}
\begin{center}
\includegraphics[width=\columnwidth, height=150mm]{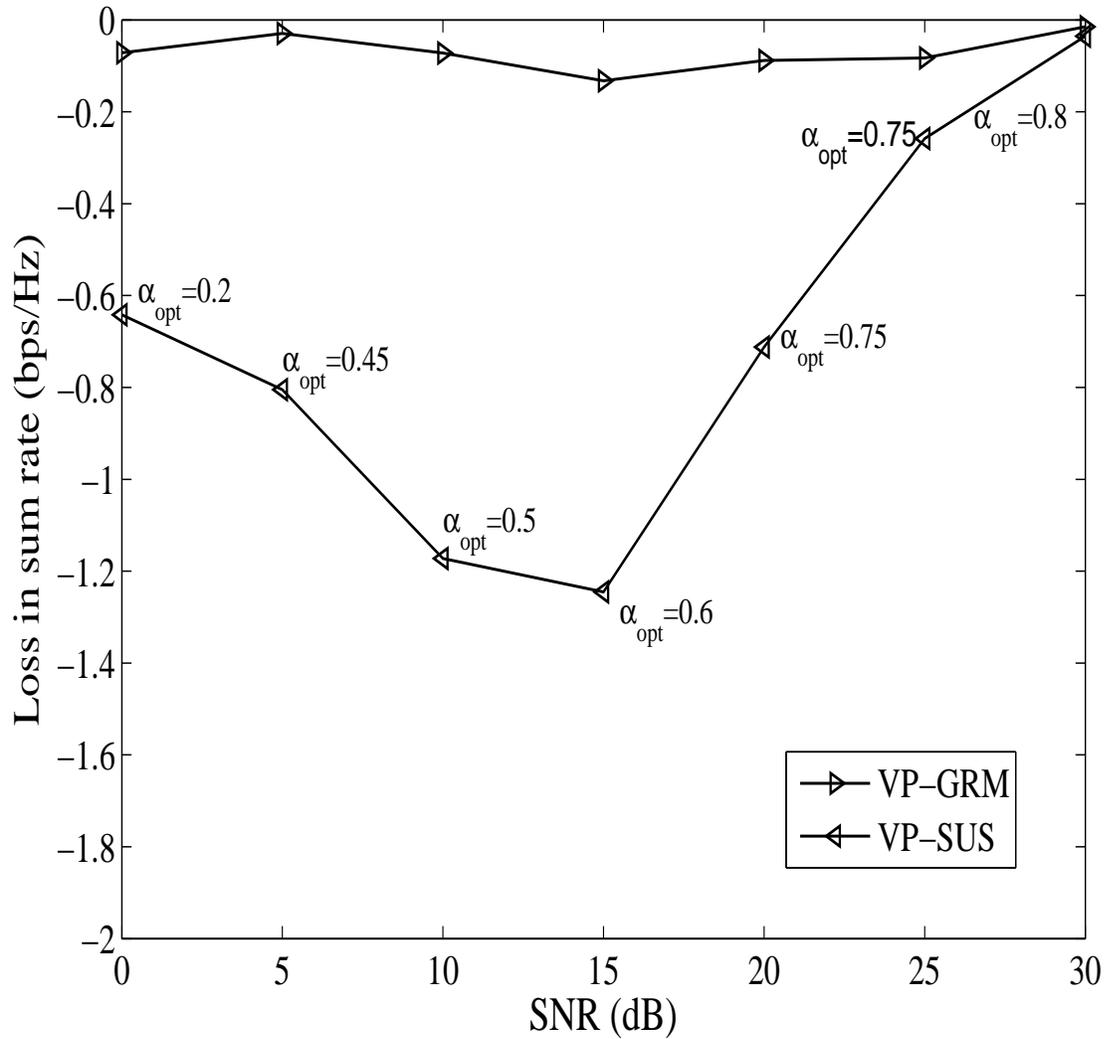}
\end{center}
\caption{Plot of loss of sum rate (bps/Hz) versus SNR (dB) for VP-GRM and VP-SUS compared to exhaustive search for VP. $N_{T} = U = 8$ and $K \leq U$.}
\label{fig:fig4}
\end{figure}

In Fig. \ref{fig:fig4}, we focus on user scheduling schemes with system parameters $N_{T}=U=8$ and $K \leq U$.  We plot the loss in sum rate of VP-GRM and VP-SUS compared to an exhaustive search for VP over all user combinations (which we denote VP-ES). Extensive simulations are used to obtain the optimal values of $\alpha$ for the VP-SUS curve, and these values are provided in the figure. We see that VP-GRM performs better than VP-SUS in the low to medium SNR region. Clearly, in this region, the GRM algorithm's sum rate based criterion is particularly effective at shedding users, compared with the SUS algorithm's orthogonality criterion. At high SNR, the two curves meet. In this region, the GRM algorithm's sum-rate based criterion is dominated by the factor $K \log (P/K)$ and thus $K=\NT$ users will always be chosen. Since the curves are on top of each other, SUS must also be choosing $K=\NT$ users, by selecting its optimal value of $\alpha$ close to $1$.

In Table \ref{table1}, we show the average number of users being selected at various SNR levels for the proposed algorithm VP-GRM and compare it with VP-SUS. We use $N_{T}=U=8$ and $K \leq U$. This table demonstrate that the two algorithms indeed perform user shedding differently. Consequently, two algorithms have different sum rate performance with VP-GRM performing better than VP-SUS.

In Table \ref{table2}, we analyze the complexity of two algorithms by averaging the total number of vector multiplications required for each algorithm. The complexity is calculated by averaging over 1000 independent channel realizations. It is obvious for GRM, we only require 2 vector multiplications in (\ref{eq:gk}), while SUS requires another vector multiplication for the user shedding operation in (\ref{eq:cos2}). However, the overall relative complexities are not obvious since the algorithms may not shed the same number of users. The table shows that the GRM complexity is in fact less than that for SUS.  The complexity of both algorithms increases with increasing SNR as they tend to shed fewer users with the increasing power levels.

\begin{figure}
\begin{center}
\includegraphics[width=\columnwidth, height=150mm]{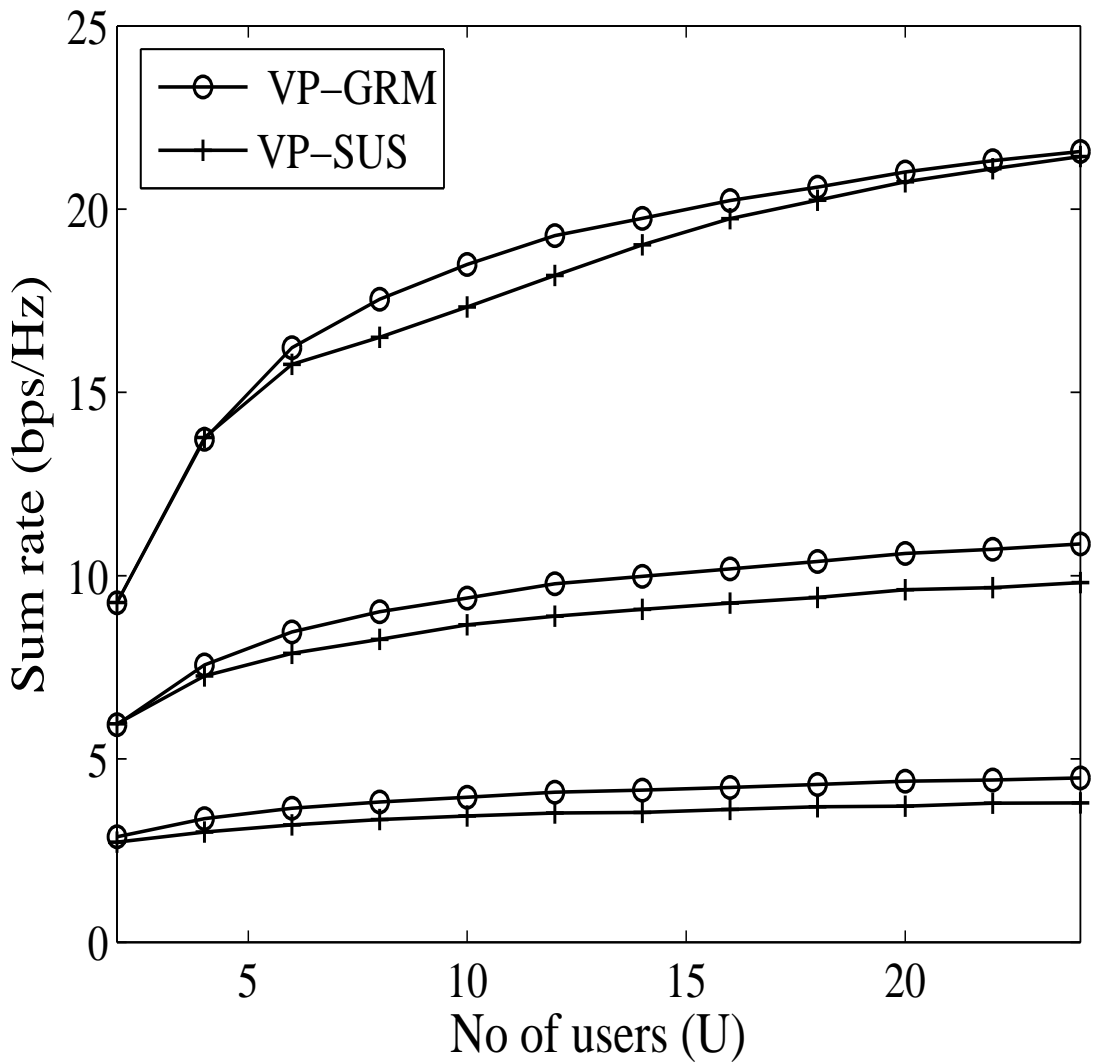}
\end{center}
\caption{ Plot of sum rate (bps/Hz) versus number of users for VP-GRM and VP-SUS. $N_T = 8$ and SNR = 0, 5 and 10 dB.}
\label{fig:fig7}
\end{figure}

In Fig. \ref{fig:fig7}, we show the performance comparison of VP-GRM and VP-SUS algorithms when $\NT= 8$ but now $U$ ranges from 2 to 24 and $K \leq \NT$. We show the sum rate results for SNR= 0, 5 and 10 dB. We again used optimal values of $\alpha$ for VP-SUS. We see that VP-GRM is performing better than VP-SUS for the whole range of $U$ for SNR = 0 and 5 dB.  But for  SNR = 10 dB, VP-SUS matches the VP-GRM performance for higher values of $U$ as both algorithms, as was discussed above, select users which are effectively in the high SNR regime hence $K$ is close to $\NT$.         

In Fig. \ref{fig:fig6}, we examine the rate allocation scheme proposed in Appendix II and the GRM based user selection algorithm. We plot the performance of the algorithms when used independently, and also for the case when the rate allocation is performed after the users are selected. We examine the scenario where $\NT$ = $U$ = 8. We see that both algorithms improve the sum rate when used independently, especially for lower SNRs. Moreover, the sum rate is barely increased when the rate allocation algorithm is applied after the user selection. This is expected from the analysis of Section \ref{sec:powerallocation}, where we see that in order to maximize the sum rate it is more important to select the users, rather than allocate (non-zero) rates to the users directly. In addition, after the user selection,  all the selected users will be operating in the high-SNR regime, and therefore there is little to be gained by performing an additional rate allocation. 

\section{Conclusion and Future Work}

In this work, we examined the sum rate of vector perturbation schemes, based on the assumptions of a uniformly distributed channel input and the tightness of the spherical Voronoi region approximation to $\Es$. We derived expressions in terms of the determinant of the channel Hermitian, and simulation results demonstrate the tightness of the bounds.

We then proceeded to the problem of individual rate allocation, as is commonly applied to other multiuser schemes to optimise the sum rate. However, we discovered that the modulo operation at the demodulator for vector perturbation precoding implies that the channel may as well be turned off when the gain is too low. Therefore only channels with high gains should be used where the energy can be applied more efficiently. Moreover, the following choice of rate allocation corresponds to standard vector perturbation precoding employing the channel inversion precoding matrix. Nevertheless, there may be a value in reconsidering the rate allocation problem with respect to scheduling fairness, different channel models, or variations of vector perturbation precoding.

It follows that user selection is the most important step to maximize the sum rate, regardless of whether the number of users exceeds that of the number of transmit antennas. Based on our high-SNR upper bound, we saw that this corresponds to determinant maximization. We proposed a greedy algorithm for this, which is essentially the same algorithm as semi-orthogonal user selection proposed in the context of ZFBF \cite{YooJSAC06}, but with more appropriate user shedding criteria, resulting in a lower-complexity and better performing algorithm which does not require optimization over the channel statistics. Naturally, the design and analysis of limited feedback techniques \cite{Love08,RyanTCPre08} for the efficient collection of CSI at the transmitter with respect to the user selection process is required. As said before, scheduling fairness among users is another important issue to consider which become all more important when all users are assumed to have same received SNR (\ie heterogeneous system model). A full treatment of this issue will be an important extension of this work in future. Also in this work, we have only considered single antenna users hence the impact of having multiple antenna receivers on the sum rate performance and scheduling complexity for vector perturbation precoding system remains an outstanding future work.  

\section*{Appendix I: Proof of Theorem \ref{th:VPcapexact}} \label{app:1}
\begin{proof}
First, note that for each $k = 1,\ldots,K$ we have
\begin{align}
I(\ah_k ; a_k)
&= H(\ah_k) - H(\ah_k|a_k).
\end{align}
Since $\ah_k$ is restricted to $\Cube$, it follows that $H(\ah_k)$ is maximized if $\ah_k$ is uniformly distributed. This is achieved if $a_k$ is uniformly distributed.
\begin{equation*}
H(\ah_k) = \log \Vol(\Cube) = 0.
\end{equation*}

In order to calculate $H(\ah_k|a_k)$, we first define few terms here. As we discussed  above, $a_k$ is uniformly distributed where we use $f(a_k)$ to denote the \text{p.d.f.} of $a_k$. Now for all $k$, we denote 
\begin{equation}
\nu_k \defas \sqbr{\eta_k}_{\mod{\Cube}},
\end{equation}
where the \text{p.d.f.} of $\nu_k$ is given by
\begin{equation}
f(\nu_k)  \defas f(\ah_k|a_k),
\end{equation}
where $f(\ah_k|a_k)$ is the \text{p.d.f.} of $\ah_k$ conditioned on $a_k$.

Noting that $f(\nu_k)$ is same for all $k$, and that $\nu_k$ is i.i.d. for the real and imaginary dimensions, we can define $\xi \defas \Real{\nu_k}$. Now, $f(\xi)$ has a modulo-Gaussian distribution given by
\begin{equation} \label{eq:fxi}
f(\xi ) \defas
\begin{cases}
\sum_{s=-\infty}^\infty \frac{1}{\sqrt{2 \pi \gamma} } e^{ - \frac{\abs{\xi - s}^2}{2 \gamma}}
& \xi \in \sqbr{-\frac{1}{2},\frac{1}{2}}, \\
0 & \xi \notin \sqbr{-\frac{1}{2},\frac{1}{2}}
\end{cases}
\end{equation}
and
\begin{equation}
\gamma  \defas \frac{\Es(\Fbf)}{2 P}.
\end{equation}

Now, to calculate $H(\ah_k|a_k)$ we have
\begin{align*}
H(\ah_k|a_k)
&= \int_{\Cube} f(a_k) \int_{\Cube} f(\ah_k| a_k ) \log f(\ah_k | a_k ) d\ah_k da_k \\
&= \int_{\Cube} f(\ah_k| a_k ) \log f(\ah_k | a_k ) d\ah_k,
\end{align*}
where the second equality follows from the fact that the inner integral is the same for all $a_k \in \Cube^K$ and that $H(\ah_k|a_k)$ is uniform. Using the definitions above, we write
\begin{equation} \label{eq:Hnu}
H(\ah_k|a_k)=H(\nu_k) = 2 H(\xi) = 2\int_{\sqbr{-\frac{1}{2},\frac{1}{2}}}  f(\xi) \log f(\xi) d\xi.
\end{equation}

Now, using $\phi(\gamma) \defas 1/\sqrt{2 \pi \gamma}$, and inserting (\ref{eq:fxi}) into (\ref{eq:Hnu}) we get
\begin{align} \label{eq:Hxi}
H(\xi) 
&~~= - \int_{-\frac{1}{2}}^{\frac{1}{2}}
\sum_{s=-\infty}^\infty \!\!\!\phi(\gamma) e^{ - \frac{\abs{\xi - s}^2}{2 \gamma}}
\log \br{
\sum_{t=-\infty}^\infty \phi(\gamma)  e^{ - \frac{\abs{\xi - t}^2}{2 \gamma}}
} d\xi \nonumber \\
&~~= \log \phi (\gamma)
- \int_{-\frac{1}{2}}^{\frac{1}{2}}
\sum_{s=-\infty}^\infty \phi(\gamma) e^{ - \frac{\abs{\xi - s}^2}{2 \gamma}}
\log {
\sum_{t=-\infty}^\infty e^{ - \frac{\abs{\xi - t}^2}{2 \gamma}}
} d\xi \nonumber \\
&~~= \frac{1}{2} \log 2 \pi e \gamma  - \Omega (\gamma)
\end{align}

where we recall the definition of $\Omega(\gamma)$ in (\ref{eq:Omegadef}). Therefore
\begin{align*}
R_{\text{VP}}(\Hbf,\Fbf) &= \sum_{k=1}^K I(\ah_k ; a_k | \Hbf, \Fbf) \\
&= - K \log \frac{\pi e \Es(\Fbf) }{P} + 2K \Omega (\gamma)\\
&= K \log \frac{P}{K} - K \log \frac{ \pi e \Es(\Fbf)}{K}+ 2K \Omega \br{\frac{\Es(\Fbf)}{2 P}}
\end{align*}
\end{proof}
which gives the theorem.

\section*{Appendix II: A Sub-optimal Rate Allocation Scheme} \label{app:2}

As we discussed in Section \ref{sec:powerallocation}, exactly solving the optimization problem of finding rate allocation matrix $\Lambdabf$ is difficult as it involve finding $\Es(\Fbf)$ which is NP-hard. Hence, we resort to a simpler sub-optimal iterative algorithm for the choice of $\Lambdabf$.

Assuming the output of each user's demodulator to be Gaussian (instead of modulo-Gaussian), the sum-rate for this $N_T \times K$ vector perturbation system is given by
\begin{equation}
R_{\text{VP-ZF}}=\sum_{k=1}^{K} \log\br{1+\delta_{k}^{2}\lambda_{k}^{2}},
\label{eq:sumrateZF}
\end{equation}
where $\delta_{k}^{2}=\frac{P}{\Es(\Fbf)}d_{k}^{2}$.

We propose to use an iterative algorithm which tries to find rate allocation matrix $\Lambdabf$ as follows:
\begin{enumerate}
\item
Initialize with lower bound on $\Es(\Fbf)$ calculated by using (\ref{eq:EsHLB}) with $\Lambdabf=\Ibf_K$
\item
Update $\Lambdabf$ by using standard waterfilling
\begin{equation}
\lambda_{k}^{2}= \max \cubr{0,\br{\zeta-\frac{1}{\delta_{k}^{2}}}},
\end{equation}
where the water level $\zeta$ is chosen as
\begin{equation}
\sum_{k=1}^{K} \max \cubr{0,\br{\zeta-\frac{1}{\delta_{k}^{2}}}}=1.
\end{equation}
\item
Update $\Es(\Fbf)$ with new precoding matrix $\Fbf$.
\item
Repeat 2) and 3) until $\Lambdabf$ converges.
\end{enumerate}
We then use this $\Lambdabf$ to calculate the sum-rate using Corollary \ref{cor:3}.  The algorithm is suboptimal because the approximation to $\Es$ is used, the received signal is assumed to be subject to Gaussian rather than modulo-Gaussian noise, and the algorithm converges to a local minimum which may not be the global minimum. 



\newpage

\newpage

\newpage
\clearpage
\newpage

\begin{table}[!h]
\caption{Average number of users selected for VP-GRM and VP-SUS. $N_{T} = U = 8$ and $K \leq U$}.%
\label{table1}%
\centering    
\begin{tabular}
[c]{|c||c|c|c|c|c|c|c|}\hline
& SNR=0dB & SNR=5dB & SNR=10dB &
SNR=15dB & SNR=20dB & SNR=25dB & SNR=30dB\\\hline
VP-GRM  & 2.3330 & 4.3480 &  6.0350 &  7.0220 & 7.5400 & 7.8370 & 7.9450 \\ \hline
VP-SUS & 2.0570 &  4.5920 &  5.4160  &  7.0480 &  7.9480 & 7.9480 & 7.9850\\\hline
\end{tabular}
\end{table}

\begin{table}[!ht]
\caption{Average number of vector multiplications for VP-GRM and VP-SUS. $N_{T} = U = 8$ and $K \leq U$.}%
\label{table2}%
\centering
\begin{tabular}
[c]{|c||c|c|c|c|}\hline
& SNR=0dB & SNR=10 dB & SNR=20dB &
SNR=30dB\\\hline
VP-GRM  & 27.4 &  62.8 &  70.8 &  71.88\\ \hline
VP-SUS & 34.5 &  64.2 &  100.8  & 104.67\\\hline
\end{tabular}
\end{table}

\newpage

\newpage
\begin{figure}
\begin{center}
\includegraphics[width=\columnwidth, height=150mm]{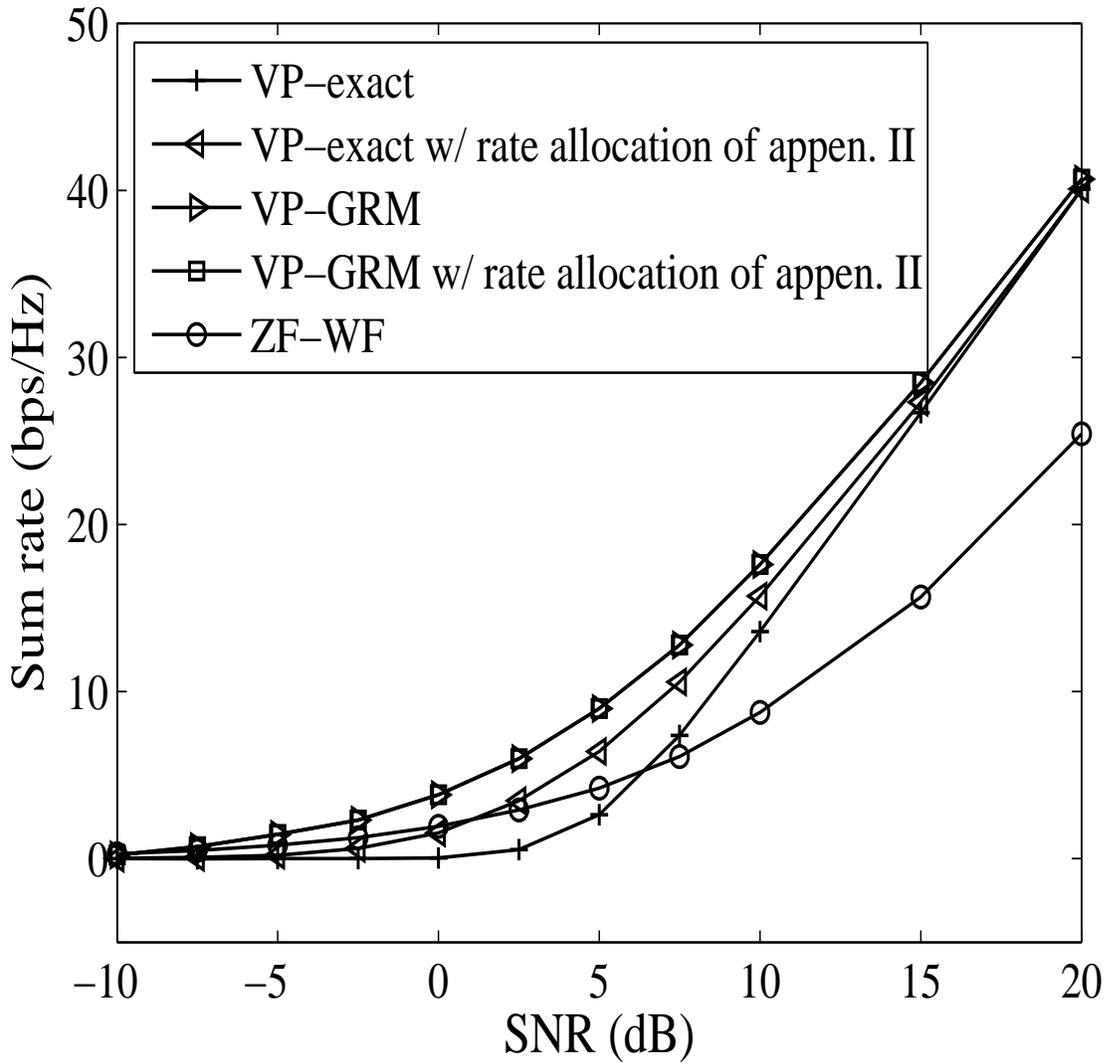}
\end{center}
\caption{Plot of sum rate (bps/Hz) versus  SNR (dB) for VP-exact, VP-exact with rate allocation from appendix II, 
VP with GRM, and VP with GRM and rate allocation from appendix II.  $N_{T} = U = 8$ and $K\leq N_{T}$.}
\label{fig:fig6}
\end{figure}

\end{document}